\documentclass[aps,prb,showkeys,floatfix,amsmath,superscriptaddress, amssymb,reprint]{revtex4-1}
\usepackage{graphicx} 
\usepackage{bm}
\usepackage{color,mathtools,bbm} 
\usepackage{wasysym}    
  \usepackage{lineno}
\usepackage{longtable}
\usepackage{footnotehyper}
\makesavenoteenv{longtable}
\usepackage{hyperref}
\usepackage{array}
\usepackage{hypernat}
\usepackage{todonotes}
\usepackage{nameref}
\usepackage{wrapfig} 
\usepackage{adjustbox}
\usepackage{multirow}
\usepackage{braket}
\usepackage{todonotes}
\begin{document}

\title{Tunable Electronic Transport in Pd$_3$O$_2$Cl$_2$ Kagome Bilayers: Interplay of Stacking Configuration and Strain}

\author{Ziao Yang}
\email{ziy226@lehigh.edu}
\affiliation{Department of Physics, Lehigh University, Bethlehem, PA 18015, USA}
\affiliation{Institute for Functional Materials \& Devices, Lehigh University, Bethlehem, PA 18015, USA}

\author{Chidiebere I. Nwaogbo}
\email{cin221@lehigh.edu}
\affiliation{Department of Physics, Lehigh University, Bethlehem, PA 18015, USA}
\affiliation{Institute for Functional Materials \& Devices, Lehigh University, Bethlehem, PA 18015, USA}

\date{\today}

\begin{abstract}
\noindent
Kagome lattice bilayers offer unique opportunities for engineering electronic properties through interlayer stacking and strain. We report a comprehensive first-principles study of Pd$_3$O$_2$Cl$_2$ kagome bilayers, examining four stacking configurations (AA, AA$'$, AB, AB$'$). Our calculations reveal dramatic stacking-dependent band gap modulation from 0.08 to 0.76~eV, with the AB$'$ configuration being the most thermodynamically stable. All stackings exhibit robust mechanical stability with Young's moduli of 54.82-61.97~N/m and ductile behavior suitable for flexible electronics. Carrier effective masses show significant stacking dependence, ranging from 2.39-6.35~$m_0$ for electrons and 0.67-1.55~$m_0$ for holes. Strain engineering of the AB$'$ bilayer demonstrates non-monotonic band gap tuning and asymmetric modulation of carrier masses, with hole effective masses showing stronger strain sensitivity. These results establish Pd$_3$O$_2$Cl$_2$ bilayers as a promising platform for strain-engineered kagome-based quantum devices, where stacking order and mechanical deformation provide complementary control over electronic transport.
\end{abstract}
  
\maketitle
\section{INTRODUCTION}
Two-dimensional (2D) materials have emerged as a transformative platform in condensed matter physics and materials science, offering tunable electronic, magnetic, and topological properties that are often inaccessible in their bulk forms \cite{chen2019electrically, zhang2016tunable,twodimensional_fan_2015}.  The reduction of dimensionality amplifies quantum confinement and many-body interactions, enabling novel physical phenomena and device applications in electronics, spintronics, and optoelectronics. Beyond graphene and transition metal dichalcogenides (TMDs), a growing focus has been on kagome lattice-based 2D systems, which host an intricate interplay between lattice geometry, orbital hybridization, and electronic correlations\cite{ye2019haas, nwaogbo2025high, li2021dirac}. The kagome lattice, composed of corner-sharing triangles, is intrinsically geometrically frustrated and is distinguished by a unique electronic band structure comprising flat bands, Dirac cones, and van Hove singularities near the Fermi level. The flat bands in particular arise from destructive interference of electronic wavefunctions, leading to strong localization and an enhanced density of states. This electronic topology provides fertile ground for correlated states such as unconventional superconductivity, magnetism, fractional quantum Hall states, and topological insulating phases.  

Recent studies have identified van der Waals kagome compounds, demonstrating that kagome networks can indeed be realized in exfoliable, and stable crystals with tunable band structures \cite{nwaogbo2025high,NWAOGBO2025101780, zhu2025theoretical, zhu2023family} . Among these materials, Pd-based kagome compounds are particularly attractive due to the alignment of Pd 4d orbitals with kagome symmetry, which facilitates the formation of ideal flat bands. In this work, Pd$_3$O$_2$Cl$_2$ exhibits semiconducting behavior with band gaps spanning the visible to ultraviolet range, as well as tunable polar properties arising from their noncentrosymmetric stacking sequences.

While monolayer kagome materials have been intensively studied, bilayer stacking introduces an additional degree of freedom that can substantially alter electronic properties. However, systematic studies of stacking-induced effects in kagome bilayers remain scarce. Interlayer coupling and stacking order can modify the orbital overlap and symmetry, thereby tuning band dispersions, effective masses, and the possible emergence of interlayer polarization. This tunability has been reported in recent advances in twisted bilayer graphene and bilayer TMDs, where stacking configuration drives correlated and topological phases \cite{ohta2006controlling, bernevig2024twisted}.

An interesting method for engineering 2D materials is through strain. External strain modifies interatomic distances and orbital hybridizations, enabling precise control of band gaps, carrier effective masses, and even semiconductor–metal transitions as demonstrated by several authors \cite{su2016band, peng2020strain}. Uniaxial strain in TMD-based heterostructures has been shown to significantly alter band alignments and induce semiconductor-to-metal transitions \cite{maesato2000uniaxial, yu2024manipulating, azizimanesh2021uniaxial}. Kagome bilayers, with their geometric sensitivity and orbital-derived flat bands, are expected to respond strongly to strain, potentially stabilizing novel electronic phases or enhancing existing correlated features,  but little attention has been given to how interlayer stacking geometry in kagome bilayers couples with strain to reshape their electronic structure. 

In this work, we address this gap by systematically investigating the strain-induced transformations in the electronic properties of Pd$_3$O$_2$Cl$_2$ bilayers under different stacking configurations (AA, AA$'$, AB and AB$'$) using first-principles density functional theory. Our study provides new insights into the interplay of geometry, orbital physics, and strain in kagome bilayers, laying the groundwork for tailoring their properties toward applications in next-generation quantum and optoelectronic devices.

\begin{figure*}[htb!]
    
    \centering
    \begin{minipage}{1.0\textwidth}
        \includegraphics[width=\textwidth]{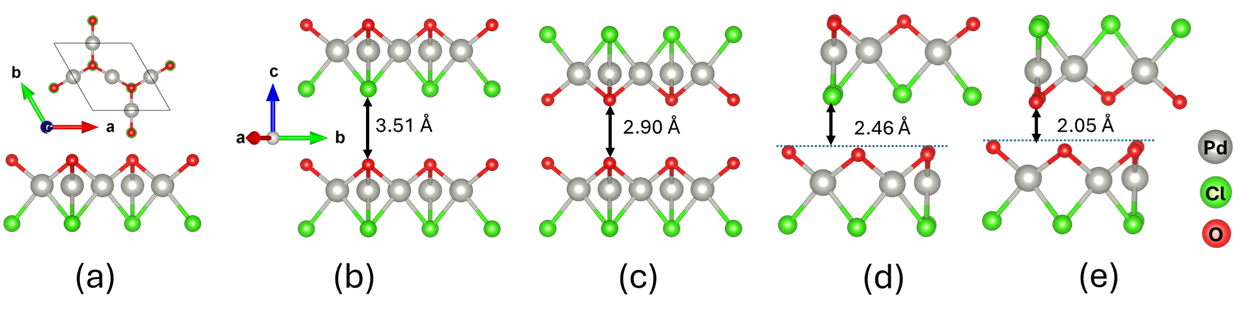}
        \caption{Crystal structures of  Pd$_3$O$_2$Cl$_2$ (a) Monolayer (top and side view) and Bilayers with different stacking configurations (b) AA  (c) AA$'$ (d) AB (e) AB$'$ }
        \label{fig1}
    \end{minipage}
\end{figure*}

\section{Computational Methods}

Bilayers (BLs) of 2D Pd$_3$O$_2$Cl$_2$ were constructed by vertically stacking monolayers of the material, to form four distinct stacking configurations AA, AA$'$, AB, and AB$'$ (see Fig. \ref{fig1} (a) - (e), while preserving the hexagonal symmetry of the monolayer. The initial interlayer distance was set to 3.5~\AA{}, and a vacuum of 35~\AA{} was applied along the $c$ axis, to minimize the artificial interactions between adjacent layers. Density functional theory (DFT) calculations were performed using the Vienna Ab initio Simulation Package (VASP) \cite{}, employing the generalized gradient approximation (GGA) with the Perdew-Burke-Ernzerhof (PBE) exchange-correlation functional~\cite{kresse1996efficiency,kresse1996efficient,perdew1996generalized}. Structural relaxation was carried out until the total energy and atomic forces converged to $\approx 10^{-8}$ eV and 0.02 eV/\AA, respectively. A $\Gamma$-centered $5 \times 5 \times 1$ k-point mesh and a plane-wave energy cutoff of 500 eV were used for these calculations. The optimized bilayer structures preserve the hexagonal symmetry, with lattice constants $a = b = 5.73$ \AA (for AA), 5.63 (AA$'$), 5.69 (AB), and 5.70 (AB$'$) and interlayer separation \textit{d} presented in Table~\ref{table1}. The relaxed monolayer structure reveals that Pd atoms preferentially occupy the corners of the lattice, forming a triangular coordination environment around the O atoms, which sit at the center of the cell with the Cl atoms as depicted in Figure \ref{fig1} (a). To evaluate the mechanical stability and probe the low-energy phonon responses, we performed 2D planar elastic constant calculations using the ElasTool toolkit~\cite{liu2022elastool,EKUMA2024109161}. For the electronic structure analysis, we employed first-principles DFT calculations.

\begin{table}[htb]
\caption{Calculated lattice constants, formation energy, interlayer distances and cleavage energies of the different stacking configurations}
\label{table1}
\renewcommand{\arraystretch}{1.1} 
\setlength{\tabcolsep}{10pt} 
\begin{tabular}{ccccccccccc}
\hline
Stacking& a & E$_f$ & \textit{d} & E$_g$ \\ \hline
AA & 5.73 &  0.069 & 3.51 & 0.08  \\
AA$'$ & 5.67 & -0.088 & 2.90 & 0.76 \\
AB  & 5.69 & -0.010 & 2.46 & 0.17  \\
AB$'$& 5.70 & -0.104 & 2.05 & 0.71 \\

\hline
\label{table1}
\end{tabular}
\end{table}

\section{RESULTS AND DISCUSSION}
\subsection{Energetics and Mechanical Properties}
\begin{table*}[htb]
 \caption{Calculated mechanical constants of bilayer $\mathrm{Pd_3O_2Cl_2}$ with different stacking configurations. 
The listed parameters include the eigenvalues of the elastic tensor ($\lambda_1$, $\lambda_2$, $\lambda_3$), planar elastic constants (C$_{11}$ and C$_{12}$), Young’s modulus (Y$^{2D}$), shear modulus (G), stiffness constant (B), Poisson’s ratio ($\nu$), average linear compressibility (C$_{avg}$), Debye temperature (T$_\mathrm{D}$), and strain energy density ($u$). 
C$_{11}$, C$_{12}$, Y$^{2D}$, G, and B are given in N/m; C$_{avg}$ in m/N; T$_\mathrm{D}$ in K; $u$ in J/m$^2$; $\lambda_{1,2,3}$,  and $\nu$ are dimensionless.
}
 \label{table2}
 \renewcommand{\arraystretch}{1.1}
 \setlength{\tabcolsep}{8pt}
 \begin{tabular}{cccccccccccccc}
 \hline
 Stacking & $\lambda_1$ & $\lambda_2$ & $\lambda_3$ & C$_{11}$ & C$_{12}$ & Y$^{2D}$ & G & B & $\nu$ & C$_{avg}$ & T$_\mathrm{D}$ & $u$ \\ \hline
 AA  & 18.727 & 37.455 & 116.657 & 77.06 & 39.60 & 56.70 & 18.73 & 58.33 & 0.51 & 0.010 & 469.60 & 0.525 \\
 AA$'$ & 20.803 & 41.605 & 121.371 & 81.49 & 39.88 & 61.97 & 20.80 & 60.69 & 0.49 & 0.0095 & 492.52 & 0.546 \\
 AB  & 17.981 & 35.962 & 115.304 & 75.63 & 39.67 & 54.82 & 17.98 & 57.65 & 0.52 & 0.01 & 461.13 & 0.519 \\
 AB$'$ & 20.719 & 41.438 & 118.745 & 80.09 & 38.65 & 61.44 & 20.72 & 59.37 & 0.48 & 0.0096 & 490.86 & 0.534 \\
 \hline
 \end{tabular}
 \end{table*}

We begin our analysis by establishing the stability of the BL configurations, using a combination of energetic, mechanical and dynamic criteria. First, we evaluated  the formation energies, using $E_{form} = E_{tot} - \sum n \epsilon_i$, where $E_{tot}$ denotes the ground-state energy of the optimized 2D Pd$_3$O$_2$Cl$_2$ BLs and $\epsilon_i$ are the ground-state energies of the constituent elements in their bulk states. The $AA$ BL showed an $E_{form}$ of 0.069 eV/atom, above the thermodynamic convex hull, while the other three configurations  ($AA'$, $AB$ and $AB'$) showed negative formation energies of -0.088 , -0.010, -0.104 ~$\text{eV}$/atom respectively (Table. \ref{table2}. These small and exothermic values indicate that the BL phases are stable relative to the elemental states. It is important to note, however, that a small positive formation energy found in the AA BL does not imply non-synthesizability. Several experimentally realized 2D materials, such as silicene and MoS$_2$, have been stabilized under nonequilibrium conditions despite positive formation energies~\cite{lalmi2010epitaxial, manni2017gdptpb, lee2012synthesis,nwaogbo2025high}. Our results suggest that similar mechanisms enable the realization of the $AA$ BL with modestly positive $E_f$ value.   Among the most stable configurations, the $AB'$ stacking exhibits the lowest formation energy ($-0.104$ eV/atom), whereas the AB stacking lies near thermodynamic neutrality, with its formation energy being closest to zero (Table~\ref{table1}). Since $AB'$  is the most thermodynamically stable configuration, we will select it as a representative material in our detailed analysis of strain-induced electronic properties.

To assess the mechanical stability and explore the low-energy lattice dynamics of the systems, we computed the two-dimensional (2D) elastic constants using the \textsc{ElasTool}. As presented in Table\ref{table2},  The resulting elastic parameters satisfy the Born stability criteria appropriate for hexagonal lattices ~\cite{born1996dynamical,PhysRevB.90.224104}, confirming that the 2D Pd$_3$O$_2$Cl$_2$ BLs are mechanically stable. The diagonalization of the elastic tensor yields eigenvalues ($\lambda_1$, $\lambda_2$, $\lambda_3$)   of 
$18.727$, $37.455$, $116.657~\text{N/m}$ for AA, $20.803$, $41.605$, $121.371~\text{N/m}$ for AA$'$, $17.981$, $35.962$, and $115.304~\text{N/m}$ for AB, and  $20.719~\text{N/m}$, $41.438~\text{N/m}$, $118.745~\text{N/m}$ for AB$'$. All eigenvalues are positive, which is a necessary and sufficient condition for mechanical stability in 2D hexagonal systems~\cite{PhysRevB.82.235414,andrew2012mechanical}. From these elastic constants, we extracted the 2D Young's modulus of $56.70$, $61.97$, $54.82$, and $61.44~\text{N/m}$ and shear modulus of $18.73$, $20.80$, $17.98$, and $20.72~\text{N/m}$ for the AA, AA$'$, AB and AB$'$ configurations, respectively, reflecting significant resistance to both tensile and shear distortions. These Young's moduli are comparable to those reported for other 2D transition metal oxides and, halides~\cite{cooper2013nonlinear,castellanos2012elastic},and  in agreement with values typical of 2D materials containing heavier transition metals such as MoS$_2$ and similar kagome-structured materials~\cite{bertolazzi2011stretching,NWAOGBO2025101780,jiang2014mechanical}. The stacking-dependent variation in Young's modulus, with AA$'$ and AB$'$ showing $\sim$12\% higher stiffness than AA and AB, indicates that interlayer arrangement significantly influences the mechanical response in BL systems. The corresponding Poisson's ratio values of $0.51$, $0.49$, $0.52$, and $0.48$ for AA, AA$'$, AB, and AB$'$, respectively, are all positive and fall within the range $0 < \nu < 1$, confirming conventional mechanical behavior with lateral contraction under uniaxial tension. Furthermore, the Pugh's modulus ratio $k = G/B$ (Table~\ref{table2}) remains below the critical threshold of $\sim$0.57~\cite{oriola2025toward, pugh1954xcii,frantsevich1982elastic}, classifying all four stacking configurations as ductile. This ductility suggests that Pd$_3$O$_2$Cl$_2$ bilayers can accommodate substantial plastic deformation before fracture, an advantageous characteristic for flexible electronics and mechanically tunable device applications. 

The dynamical stability was further examined through the acoustic phonon responses of the materials.  The sound velocities extracted from the slopes of the longitudinal and transverse acoustic modes near the $\Gamma$ point show a clear stacking dependence. For the AA configuration, the longitudinal and transverse velocities are $V_l = 3.96~\text{km/s}$ and $V_s = 1.95~\text{km/s}$, respectively, with an average velocity of $V_d = 2.47~\text{km/s}$. In the AA$'$ stacking, the corresponding values increase slightly to $V_l = 4.06~\text{km/s}$, $V_s = 2.05~\text{km/s}$, and $V_d = 2.59~\text{km/s}$. For the AB configuration, the extracted velocities are $V_l = 3.92~\text{km/s}$, $V_s = 1.91~\text{km/s}$, and $V_d = 2.43~\text{km/s}$. Finally, the AB$'$ stacking exhibits velocities of $V_l = 4.03~\text{km/s}$, $V_s = 2.05~\text{km/s}$, and $V_d = 2.58~\text{km/s}$. These velocities are in direct correspondence with the elastic stiffness and characterize the propagation of mechanical perturbations through the 2D lattice \cite{born1996dynamical, jiang2014mechanical}. The estimated Debye temperatures also show a clear stacking dependence. For the AA configuration, the Debye temperature is $T_D = 469.60~\text{K}$. In the AA$'$ stacking, it increases to $T_D = 492.52~\text{K}$. For the AB configuration, the value slightly decreases to $T_D = 461.13~\text{K}$, while the AB$'$ stacking yields a higher value of $T_D = 490.86~\text{K}$. The magnitude of these Debye temperatures further attests to the robustness of the phonon spectrum and the rigidity of the low-energy vibrations across all stacking configurations.
 Collectively, the favorable elastic constants, relatively high sound velocities, and elevated Debye temperature signify excellent mechanical and dynamical stability, an essential prerequisite for the practical realization of 2D Pd$_3$O$_2$Cl$_2$ BLs in flexible and resilient quantum material architectures.

\begin{figure}[t!]
	\centering
	\includegraphics[trim = 0mm 0mm 0mm 0mm,width=\linewidth,clip=true]{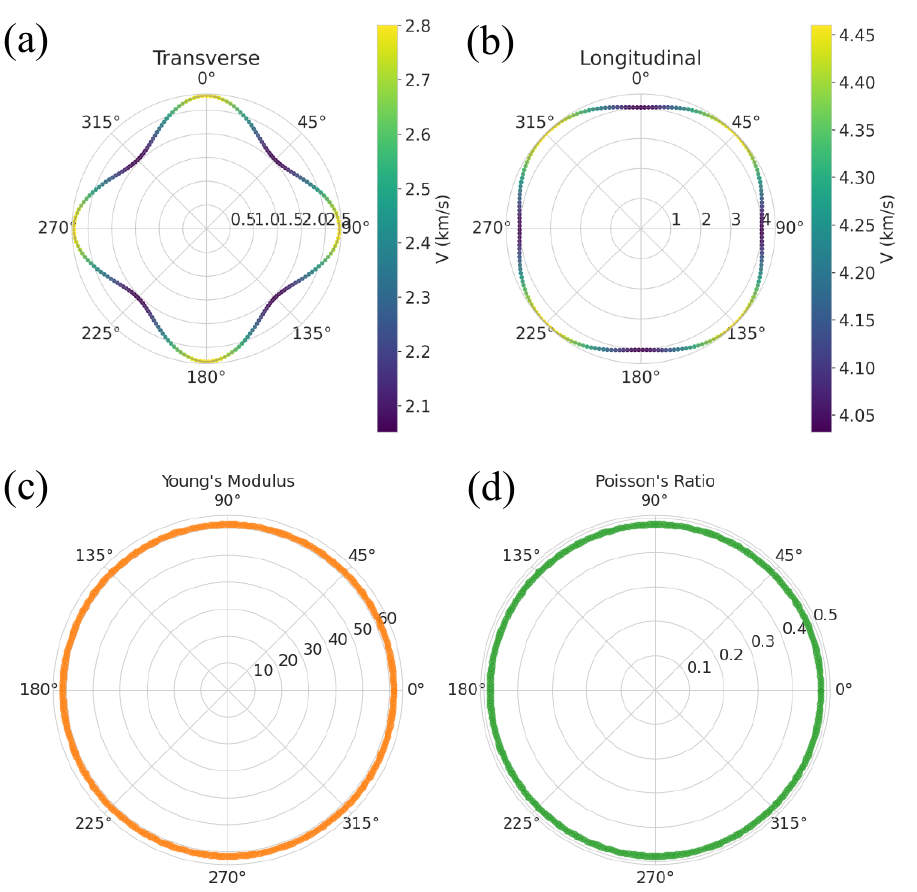} 
	\caption{
Angular dependence of in–plane acoustic velocities and elastic moduli for the AB$'$ Pd$_3$O$_2$Cl$_2$ bilayer.
(a) Polar plot of the transverse acoustic sound velocity $V_\mathrm{T}(\theta)$, showing a weak four–lobed modulation associated with slight shear–wave anisotropy.
(b) Polar plot of the longitudinal acoustic sound velocity $V_\mathrm{L}(\theta)$, which is nearly isotropic and varies only weakly with direction.
(c) Directional Young’s modulus $E(\theta)$, exhibiting an almost perfectly circular contour and indicating quasi–isotropic in–plane stiffness.
(d) Directional Poisson’s ratio $\nu(\theta)$, which remains positive and nearly constant for all in–plane directions, confirming the absence of auxetic behavior and the overall elastic isotropy of the AB$'$ bilayer.
}
 
	\label{fig2a}
\end{figure}

Figure~\ref{fig2a}(a-d) summarizes the in-plane elastic  and acoustic anisotropy of the most stable AB$'$ Pd$_3$O$_2$Cl$_2$ BL. In Figs.~\ref{fig2a}(a) and (b) we plot the angular dependence of the transverse and longitudinal acoustic sound velocities, $V_\mathrm{T}(\theta)$ and $V_\mathrm{L}(\theta)$, for in–plane propagation.  For the transverse branch (Fig.~\ref{fig2a}(a)), the polar contour exhibits a clear fourfold modulation, with the shear wave velocity $V_\mathrm{T}$ varying between $\sim 2.1$ and $\sim 2.8$~km/s as the propagation direction is rotated in the plane. The maxima occur along the mechanically stiffer directions, while the minima coincide with the softer axes of the kagome network, reflecting the underlying anisotropy of the in-plane shear response. In contrast, the longitudinal acoustic mode (Fig.~\ref{fig2a}(b)) is nearly isotropic: the longitudinal velocity $V_\mathrm{L}$ remains confined to a narrow window,  producing an almost circular polar distribution with only a weak fourfold distortion. This indicates that the bulk (compressional) stiffness of the AB$'$ bilayer is essentially direction independent, while the anisotropy is primarily carried by the shear channel. The modest degree of acoustic anisotropy, together with the absence of vanishing velocities in any direction, suggests mechanically robust in-plane wave propagation and supports the use of continuum elastic descriptions for long-wavelength phonons in this kagome bilayer. Figs \ref{fig2a} (c) and (d) show the corresponding directional Young’s modulus $E(\theta)$ and Poisson’s ratio $\nu(\theta)$ extracted from the 2D elastic tensor. Both polar plots are essentially circular: $E(\theta)$ remains nearly constant at $\sim 55$–$60$~N/m for all in–plane orientations, while $\nu(\theta)$ hovers around $\sim 0.45$–$0.50$ with only minor angular variation. This indicates that the AB$'$ bilayer behaves as an almost elastically isotropic membrane in the linear regime, with anisotropy ratios $A_E$ and $A_\nu$ very close to unity. The relatively large and positive Poisson’s ratio reflects strong lateral contraction under uniaxial tension. Together with the strain–tunable band structure discussed below, this elastic profile suggests that AB$'$ Pd$_3$O$_2$Cl$_2$ can accommodate external deformation without developing strong direction–dependent soft modes, making it a promising platform for flexible and strain–engineered kagome–based devices.
\vspace{-1.5em}

\begin{figure*}[htb!]
    
    \centering
    \begin{minipage}{1.0\textwidth}
        \includegraphics[width=\textwidth]{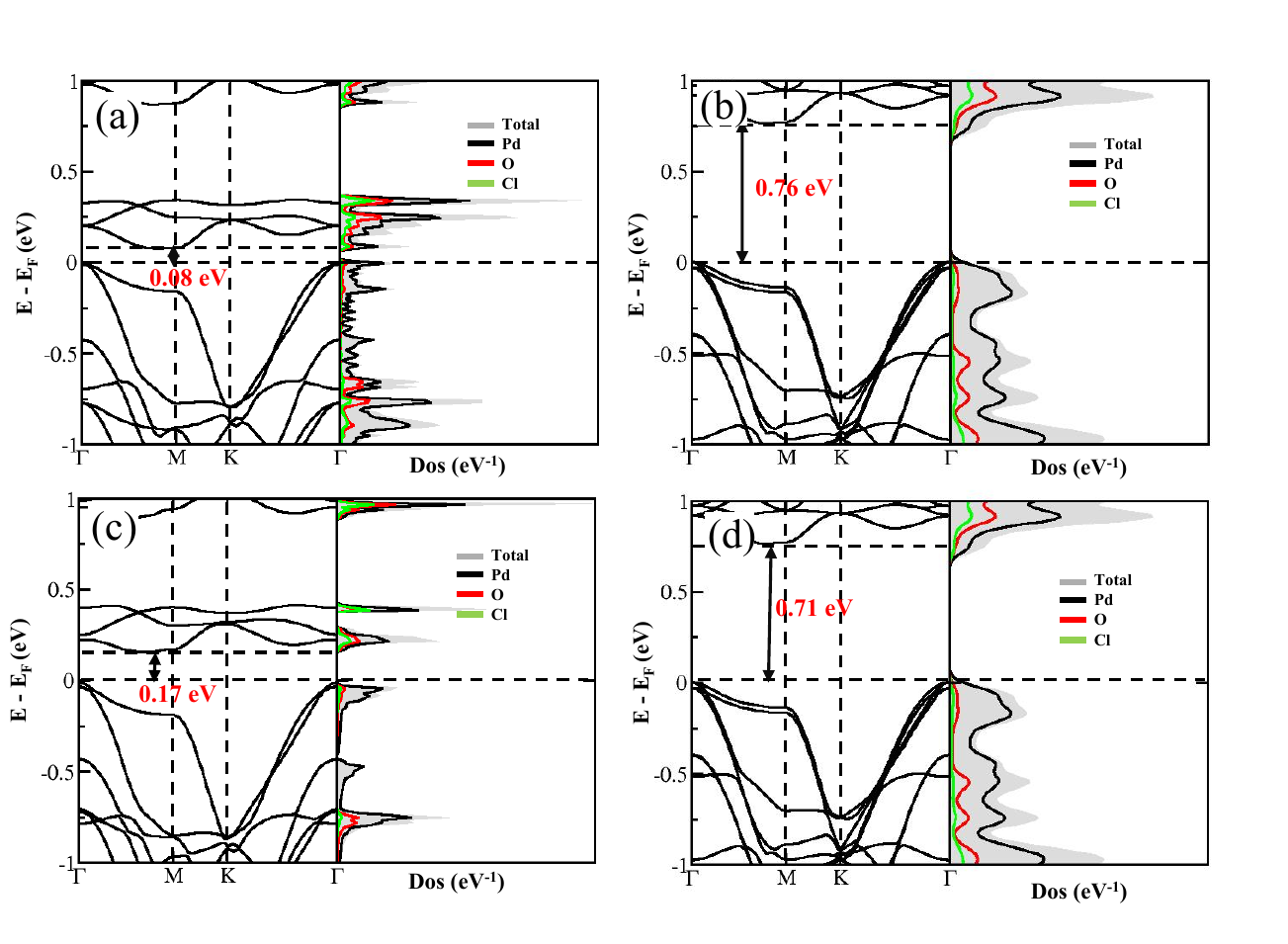}
        \caption{Band structures and density of states of BL  Pd$_3$O$_2$Cl$_2$ (a) $AA$  (b) $AA'$ (c) $AB$ (d) $AB'$. High-registry stackings (AA and AB) show narrow semiconducting gaps of 0.08 eV and
    0.17 eV, respectively, due to enhanced Pd–4$d$ interlayer orbital overlap, while the flipped
    and shifted configurations (AA$'$ and AB$'$) yield substantially larger gaps of 0.76 eV and
    0.71 eV. The DOS highlights the dominant contributions of Pd~4$d$ and O~2$p$ orbitals
    near the valence band, and Pd~4$d$/Cl~$p$ hybridization at the conduction edge, illustrating
    how stacking order governs the redistribution of spectral weight and gap opening in the
    BL system. }
        \label{fig2}
    \end{minipage}
\end{figure*}

\subsection{Electronic Properties}
Figure~\ref{fig2} presents the stacking–dependent electronic structure of Pd$_3$O$_2$Cl$_2$ BLs, showing the band dispersions and projected density of states (DOS) for the four configurations AA, AA$'$, AB, and AB$'$. In all cases the characteristic kagome behaviors are preserved, namely dispersive bands crossing near the K point and  relatively flat bands in the vicinity of the Fermi level, consistent with earlier studies of kagome oxohalides and related Pd–based kagome systems~\cite{zhou2014sd,yin2019negative,park2020kagome,li2025electronic}. However, the interlayer stacking exerts a pronounced control over the size and nature of the band gap. The AA and AB stackings remain semiconducting but with very small gaps of $E_\mathrm{g}\!\approx\!0.08$~eV and $0.17$~eV, respectively, where the valence and conduction band edges lie close in energy along the $\Gamma$–K path. This behavior reflects enhanced interlayer Pd–4$d$ hybridization in high–registry stackings, which brings the kagome–derived conduction and valence manifolds closer to the Fermi level and partially narrows the gap. By contrast, the AA$'$ and AB$'$ stackings, where one layer is flipped and laterally shifted  with respect to the other, exhibit substantially larger gaps of $E_\mathrm{g}\!\approx\!0.76$~eV and $0.71$~eV. Here the reduced direct Pd-Pd overlap and modified Pd-O-Pd hopping pathways weaken the interlayer coupling, leading to a more pronounced bonding–antibonding splitting of the band-edge states and thus a wider gap. This strong sensitivity of the band gap to stacking geometry is consistent with the general picture established in bilayer graphene, TMD bilayers, and other van der Waals heterostructures, where layer alignment and registry tune both band dispersions and gap size~\cite{cusati2018stacking, ma2014tunable, hu2025tunable, yun2018tunable,ohta2006controlling}. The accompanying total and projected DOS further reinforces this picture. In every stacking, the states near the valence–band maximum are dominated by hybridized Pd~4$d$ and O~2$p$ orbitals, while the conduction–band minimum primarily involves Pd~4$d$ states with appreciable Cl~$p$ contributions, highlighting the crucial role of $d$–$p$ covalency in shaping the kagome bands. For the small–gap AA and AB configurations, the DOS shows sizeable spectral weight extending towards the Fermi level, with sharp Pd–$d$–derived peaks just below and above $E_\mathrm{F}$ that originate from nearly flat kagome bands and van Hove singularities. In contrast, the AA$'$ and AB$'$ stackings display a much cleaner depletion of the DOS around $E_\mathrm{F}$, reflecting their larger band gaps, and sharper DOS peaks at the band edges, indicative of stronger localization of the kagome-derived states.

To quantify how different stacking configurations influence carrier transport, we extracted the electron and hole effective masses for each structure by fitting the band edges to parabolic dispersions. The electron effective masses are $6.35$, $4.92$, $4.25$, and $2.39~m_0$ for the AA, AA$'$, AB, and AB$'$ configurations, respectively. The corresponding hole effective masses are $1.55$, $1.53$, $0.95$, and $0.67~m_0$ for the AA, AA$'$, AB, and AB$'$ configurations, respectively. These values exhibit a pronounced stacking dependence, with the AB$'$ configuration showing the lightest carriers for both electrons and holes, consistent with its wider band gap and enhanced interlayer orbital hybridization. The calculated effective masses place Pd$_3$O$_2$Cl$_2$ BLs in an intermediate regime between highly anisotropic 2D materials and more isotropic systems. For comparison, BL graphene exhibits electron and hole effective masses of 0.03–0.05 $m_0$~\cite{zou2011effective}, while bilayer MoS$_2$ shows significantly heavier carriers with electron masses of $\sim$0.48–0.60 $m_0$ and hole masses of $\sim$0.54–0.62 $m_0$ depending on stacking configuration~\cite{he2014stacking}. The 2D perovskite (PEA)$_2$PbI$_4$ displays comparable effective masses of 0.45-0.76$~m_0$~\cite{ma2017layer}. The stacking-induced modulation of effective masses in Pd$_3$O$_2$Cl$_2$ is comparable to that reported for twisted bilayer MoS$_2$, where twist angle variations can tune effective masses by factors of 2-3~\cite{tan2016first,yeh2018direct}.

\subsection{Strain-induced electronic properties}

\begin{figure}[htb!]
	\centering
	\includegraphics[trim = 0mm 0mm 0mm 0mm,width=\linewidth,clip=true]{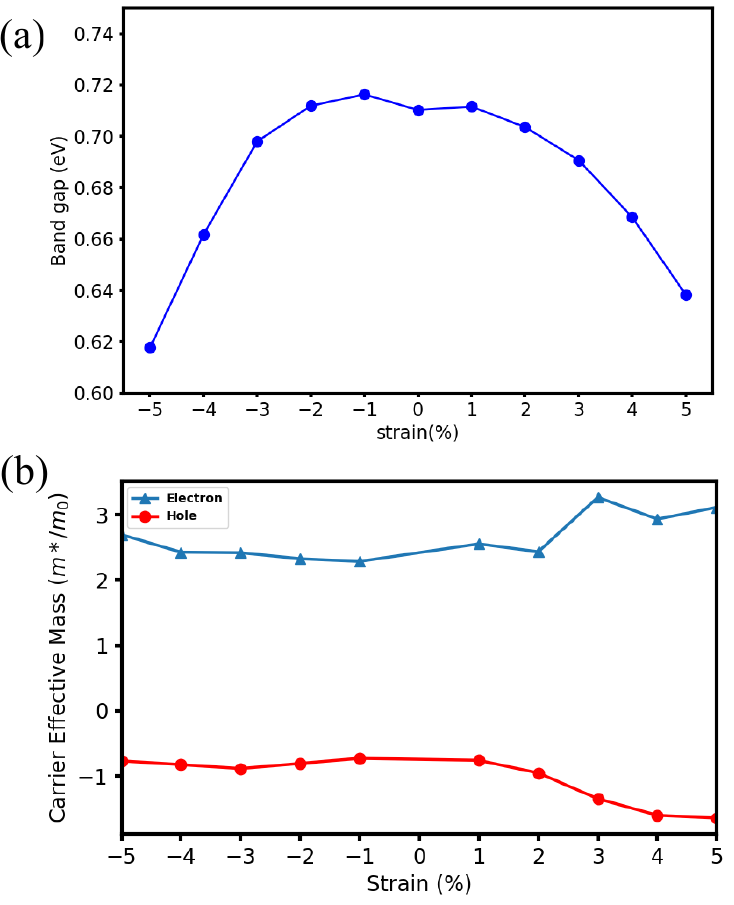} 
	\caption{Strain induced effect on the AB' BL (a) Band gap as a function of strain. (b) Electron and hole effective masses under uniaxial strain.} 
	\label{fig4}
\end{figure}

Figure~\ref{fig4} summarizes the strain–tunable electronic response of the thermodynamically most stable AB$'$ bilayer, as identified from the formation energies $E_{\mathrm{f}}$ in Table~\ref{table1}. Motivated by its energetic stability and favorable band gap at zero strain, we apply uniaxial strain $\varepsilon$ along the in–plane $a$ direction, ranging from $\varepsilon=-5\%$ (compressive) to $\varepsilon=+5\%$ (tensile) in steps of $1\%$, and for each configuration fully relax the internal atomic coordinates before computing the electronic structure. The resulting evolution of the band gap $E_{\mathrm{g}}(\varepsilon)$ is shown in Fig.~\ref{fig4}(a). For small $|\varepsilon|$ the band gap is slightly enhanced relative to its equilibrium value, exhibiting a shallow maximum under modest compressive strain. This reflects an initial optimization of the bonding–antibonding splitting between the kagome-derived valence and conduction–band edges when the Pd–O–Pd bond lengths and angles are gently shortened along the strain direction, which increases the in–plane hopping amplitude and pushes the band extrema slightly apart. As the magnitude of the applied strain is further increased, the gap decreases for both compressive and tensile loading, indicating that stronger uniaxial distortion eventually overwhelms this optimization and drives the valence and conduction manifolds closer together. Physically, large compressive strain enhances the anisotropy of the kagome network and can shift the band extrema away from their unstrained $k$–points, while tensile strain weakens the in–plane Pd-O hybridization and reduces the bonding–antibonding splitting, in both cases narrowing the gap. This non–monotonic response of $E_{\mathrm{g}}(\varepsilon)$, with an initial increase followed by a pronounced reduction at larger $|\varepsilon|$, is consistent with the general strain–engineering picture established in other two–dimensional semiconductors and van der Waals materials, where moderate strain can open or enlarge band gaps, whereas stronger strain tends to reduce them and may even induce indirect–gap or semimetallic behavior\cite{cusati2018stacking, ma2014tunable, yun2018tunable}.

To quantify how such lattice distortions impact carrier transport, we also extract the effective masses of electrons and holes for each strained structure, as plotted in Fig.~\ref{fig4}(b). The effective masses $m^{\ast}$ are obtained by sampling the band dispersion in the vicinity of the conduction–band minimum (CBM) and valence–band maximum (VBM) along the transport direction, fitting $E(k)$ to a quadratic form, and evaluating the curvature via the standard parabolic approximation,  $m^{\ast} = \frac{\hbar^{2}}{\mathrm{d}^{2}E/\mathrm{d}k^{2}} \Big|_{k=k_{\mathrm{ext}}}.$
We find that the electron effective mass $m^{\ast}_{e}$ varies only weakly over the entire strain window, indicating that the conduction band retains a similar curvature and bandwidth even under relatively strong uniaxial distortion. In contrast, the hole effective mass $m^{\ast}_{h}$ exhibits a much stronger strain dependence, particularly under tensile strain, where the VBM becomes noticeably flatter along the strained direction. This pronounced renormalization of $m^{\ast}_{h}$ reflects the greater sensitivity of the valence–band states, dominated by more localized Pd~4$d$ and O~2$p$ orbitals in the kagome network to changes in bond angles and local crystal field, whereas the more delocalized conduction–band states respond more gently to the same structural perturbations. The resulting disparity between the strain responses of $m^{\ast}_{e}$ and $m^{\ast}_{h}$ implies that uniaxial strain provides an efficient knob to selectively tune hole mobility while leaving the electron channel comparatively robust, a behavior that parallels previous reports in strained 2D semiconductors where hole effective masses are strongly modulated by relatively small applied strains\cite{yun2018tunable, tan2016first, yeh2018direct}. Overall, Fig.~\ref{fig4} demonstrates that the AB$'$ Pd$_3$O$_2$Cl$_2$ bilayer combines thermodynamic stability with a highly strain–responsive band gap and hole effective mass, making it a promising platform for strain-engineered kagome based quantum devices.

\section{CONCLUSION}
We have carried out a first-principles investigation of the kagome bilayer Pd$_3$O$_2$Cl$_2$ and shown that both stacking configuration and external strain play central roles in governing its structural and electronic behavior. For all four considered stackings - AA, AA$'$, AB, and AB$'$, we computed the electronic band structures and densities of states, revealing how changes in interlayer configurations reshape the band dispersion and spectral weight near the Fermi level. By comparing formation energies, we identified the AB$'$ arrangement as the most stable stacking, and subsequently selected it as the reference structure for a more detailed analysis of strain-induced and mechanical responses. In this phase, our calculations demonstrate that the band gap can be continuously tuned by applied strain, accompanied by pronounced changes in the effective masses at the band edges, indicating a strong sensitivity of carrier curvature and transport characteristics to external deformation. Together with the elastic properties obtained for the AB$'$ bilayer, these results highlight Pd$_3$O$_2$Cl$_2$ as a mechanically robust yet electronically flexible kagome system, where stacking order and strain act as powerful knobs for band engineering. This positions Pd$_3$O$_2$Cl$_2$ as a promising platform for exploring structure-property coupling and mechanically controlled electronic functionalities in $d$-orbital dominated two-dimensional kagome materials.

\vspace{0.2cm}
\noindent \textbf{ACKNOWLEDGEMENT}

This research is supported by the U.S. Department of Energy, Office of Science, Basic Energy Sciences under Award DOE-SC0024099. This work used Stampede3 at the Texas Advanced Computing Center (TACC) through allocation PHY240252 from the Advanced Cyberinfrastructure Coordination Ecosystem: Services \& Support (ACCESS) program, supported by the NSF Grants 2138259, 2138286, 2138307, 2137603, and 2138296. Computational resources were also provided by Lehigh University's Research Computing Infrastructure.

\vspace{0.2cm}
\noindent \textbf{Author Contributions:} \textbf {Z. Yang} Data Curation, Formal Analysis, Investigation, Writing-Original Draft, Visualization, Investigation, Methodology, and Writing-Review \& Editing. \textbf{C. I. Nwaogbo} Conceptualization, Supervision, Formal Analysis, Writing-Original Draft, Visualization, validation, Investigation, Methodology, Writing-Review \& Editing. 
\vspace{0.2cm}

\noindent \textbf{Declaration of Competing Interest}
\noindent The authors declare that they have no known competing financial interests or personal relationships that could have appeared to influence the work reported in this paper.


\begin{thebibliography}{48}%
\makeatletter
\providecommand \@ifxundefined [1]{%
 \@ifx{#1\undefined}
}%
\providecommand \@ifnum [1]{%
 \ifnum #1\expandafter \@firstoftwo
 \else \expandafter \@secondoftwo
 \fi
}%
\providecommand \@ifx [1]{%
 \ifx #1\expandafter \@firstoftwo
 \else \expandafter \@secondoftwo
 \fi
}%
\providecommand \natexlab [1]{#1}%
\providecommand \enquote  [1]{``#1''}%
\providecommand \bibnamefont  [1]{#1}%
\providecommand \bibfnamefont [1]{#1}%
\providecommand \citenamefont [1]{#1}%
\providecommand \href@noop [0]{\@secondoftwo}%
\providecommand \href [0]{\begingroup \@sanitize@url \@href}%
\providecommand \@href[1]{\@@startlink{#1}\@@href}%
\providecommand \@@href[1]{\endgroup#1\@@endlink}%
\providecommand \@sanitize@url [0]{\catcode `\\12\catcode `\$12\catcode `\&12\catcode `\#12\catcode `\^12\catcode `\_12\catcode `\%12\relax}%
\providecommand \@@startlink[1]{}%
\providecommand \@@endlink[0]{}%
\providecommand \url  [0]{\begingroup\@sanitize@url \@url }%
\providecommand \@url [1]{\endgroup\@href {#1}{\urlprefix }}%
\providecommand \urlprefix  [0]{URL }%
\providecommand \Eprint [0]{\href }%
\providecommand \doibase [0]{http://dx.doi.org/}%
\providecommand \selectlanguage [0]{\@gobble}%
\providecommand \bibinfo  [0]{\@secondoftwo}%
\providecommand \bibfield  [0]{\@secondoftwo}%
\providecommand \translation [1]{[#1]}%
\providecommand \BibitemOpen [0]{}%
\providecommand \bibitemStop [0]{}%
\providecommand \bibitemNoStop [0]{.\EOS\space}%
\providecommand \EOS [0]{\spacefactor3000\relax}%
\providecommand \BibitemShut  [1]{\csname bibitem#1\endcsname}%
\let\auto@bib@innerbib\@empty
\bibitem [{\citenamefont {Chen}\ \emph {et~al.}(2019)\citenamefont {Chen}, \citenamefont {Zhou}, \citenamefont {Deng}, \citenamefont {Wu}, \citenamefont {Xia}, \citenamefont {Cao}, \citenamefont {Zhang}, \citenamefont {Huang}, \citenamefont {Wang},\ and\ \citenamefont {Wang}}]{chen2019electrically}%
  \BibitemOpen
  \bibfield  {author} {\bibinfo {author} {\bibfnamefont {X.}~\bibnamefont {Chen}}, \bibinfo {author} {\bibfnamefont {Z.}~\bibnamefont {Zhou}}, \bibinfo {author} {\bibfnamefont {B.}~\bibnamefont {Deng}}, \bibinfo {author} {\bibfnamefont {Z.}~\bibnamefont {Wu}}, \bibinfo {author} {\bibfnamefont {F.}~\bibnamefont {Xia}}, \bibinfo {author} {\bibfnamefont {Y.}~\bibnamefont {Cao}}, \bibinfo {author} {\bibfnamefont {L.}~\bibnamefont {Zhang}}, \bibinfo {author} {\bibfnamefont {W.}~\bibnamefont {Huang}}, \bibinfo {author} {\bibfnamefont {N.}~\bibnamefont {Wang}}, \ and\ \bibinfo {author} {\bibfnamefont {L.}~\bibnamefont {Wang}},\ }\href@noop {} {\bibfield  {journal} {\bibinfo  {journal} {Nano Today}\ }\textbf {\bibinfo {volume} {27}},\ \bibinfo {pages} {99} (\bibinfo {year} {2019})}\BibitemShut {NoStop}%
\bibitem [{\citenamefont {Zhang}\ \emph {et~al.}(2016)\citenamefont {Zhang}, \citenamefont {Liu}, \citenamefont {Yu}, \citenamefont {Hang}, \citenamefont {Li}, \citenamefont {Guo}, \citenamefont {Xu}, \citenamefont {Sun}, \citenamefont {Zhou},\ and\ \citenamefont {Guo}}]{zhang2016tunable}%
  \BibitemOpen
  \bibfield  {author} {\bibinfo {author} {\bibfnamefont {Z.}~\bibnamefont {Zhang}}, \bibinfo {author} {\bibfnamefont {X.}~\bibnamefont {Liu}}, \bibinfo {author} {\bibfnamefont {J.}~\bibnamefont {Yu}}, \bibinfo {author} {\bibfnamefont {Y.}~\bibnamefont {Hang}}, \bibinfo {author} {\bibfnamefont {Y.}~\bibnamefont {Li}}, \bibinfo {author} {\bibfnamefont {Y.}~\bibnamefont {Guo}}, \bibinfo {author} {\bibfnamefont {Y.}~\bibnamefont {Xu}}, \bibinfo {author} {\bibfnamefont {X.}~\bibnamefont {Sun}}, \bibinfo {author} {\bibfnamefont {J.}~\bibnamefont {Zhou}}, \ and\ \bibinfo {author} {\bibfnamefont {W.}~\bibnamefont {Guo}},\ }\href@noop {} {\bibfield  {journal} {\bibinfo  {journal} {Wiley Interdisciplinary Reviews: Computational Molecular Science}\ }\textbf {\bibinfo {volume} {6}},\ \bibinfo {pages} {324} (\bibinfo {year} {2016})}\BibitemShut {NoStop}%
\bibitem [{\citenamefont {Fan}\ \emph {et~al.}(2015)\citenamefont {Fan}, \citenamefont {Li},\ and\ \citenamefont {Djerdj}}]{twodimensional_fan_2015}%
  \BibitemOpen
  \bibfield  {author} {\bibinfo {author} {\bibfnamefont {J.}~\bibnamefont {Fan}}, \bibinfo {author} {\bibfnamefont {T.}~\bibnamefont {Li}}, \ and\ \bibinfo {author} {\bibfnamefont {I.}~\bibnamefont {Djerdj}},\ }\href {\doibase 10.1007/s11664-015-3947-6} {\bibfield  {journal} {\bibinfo  {journal} {Journal of Electronic Materials}\ } (\bibinfo {year} {2015}),\ 10.1007/s11664-015-3947-6}\BibitemShut {NoStop}%
\bibitem [{\citenamefont {Ye}\ \emph {et~al.}(2019)\citenamefont {Ye}, \citenamefont {Chan}, \citenamefont {McDonald}, \citenamefont {Graf}, \citenamefont {Kang}, \citenamefont {Liu}, \citenamefont {Suzuki}, \citenamefont {Comin}, \citenamefont {Fu},\ and\ \citenamefont {Checkelsky}}]{ye2019haas}%
  \BibitemOpen
  \bibfield  {author} {\bibinfo {author} {\bibfnamefont {L.}~\bibnamefont {Ye}}, \bibinfo {author} {\bibfnamefont {M.~K.}\ \bibnamefont {Chan}}, \bibinfo {author} {\bibfnamefont {R.~D.}\ \bibnamefont {McDonald}}, \bibinfo {author} {\bibfnamefont {D.}~\bibnamefont {Graf}}, \bibinfo {author} {\bibfnamefont {M.}~\bibnamefont {Kang}}, \bibinfo {author} {\bibfnamefont {J.}~\bibnamefont {Liu}}, \bibinfo {author} {\bibfnamefont {T.}~\bibnamefont {Suzuki}}, \bibinfo {author} {\bibfnamefont {R.}~\bibnamefont {Comin}}, \bibinfo {author} {\bibfnamefont {L.}~\bibnamefont {Fu}}, \ and\ \bibinfo {author} {\bibfnamefont {J.~G.}\ \bibnamefont {Checkelsky}},\ }\href@noop {} {\bibfield  {journal} {\bibinfo  {journal} {Nature communications}\ }\textbf {\bibinfo {volume} {10}},\ \bibinfo {pages} {4870} (\bibinfo {year} {2019})}\BibitemShut {NoStop}%
\bibitem [{\citenamefont {Nwaogbo}\ \emph {et~al.}(2025{\natexlab{a}})\citenamefont {Nwaogbo}, \citenamefont {Iloanya},\ and\ \citenamefont {Ekuma}}]{nwaogbo2025high}%
  \BibitemOpen
  \bibfield  {author} {\bibinfo {author} {\bibfnamefont {C.~I.}\ \bibnamefont {Nwaogbo}}, \bibinfo {author} {\bibfnamefont {A.~C.}\ \bibnamefont {Iloanya}}, \ and\ \bibinfo {author} {\bibfnamefont {C.~E.}\ \bibnamefont {Ekuma}},\ }\href@noop {} {\bibfield  {journal} {\bibinfo  {journal} {Computational Materials Science}\ }\textbf {\bibinfo {volume} {259}},\ \bibinfo {pages} {114125} (\bibinfo {year} {2025}{\natexlab{a}})}\BibitemShut {NoStop}%
\bibitem [{\citenamefont {Li}\ \emph {et~al.}(2021)\citenamefont {Li}, \citenamefont {Wang}, \citenamefont {Wang}, \citenamefont {Yuan}, \citenamefont {Song}, \citenamefont {Lou}, \citenamefont {Liu}, \citenamefont {Huang}, \citenamefont {Liu}, \citenamefont {Lei} \emph {et~al.}}]{li2021dirac}%
  \BibitemOpen
  \bibfield  {author} {\bibinfo {author} {\bibfnamefont {M.}~\bibnamefont {Li}}, \bibinfo {author} {\bibfnamefont {Q.}~\bibnamefont {Wang}}, \bibinfo {author} {\bibfnamefont {G.}~\bibnamefont {Wang}}, \bibinfo {author} {\bibfnamefont {Z.}~\bibnamefont {Yuan}}, \bibinfo {author} {\bibfnamefont {W.}~\bibnamefont {Song}}, \bibinfo {author} {\bibfnamefont {R.}~\bibnamefont {Lou}}, \bibinfo {author} {\bibfnamefont {Z.}~\bibnamefont {Liu}}, \bibinfo {author} {\bibfnamefont {Y.}~\bibnamefont {Huang}}, \bibinfo {author} {\bibfnamefont {Z.}~\bibnamefont {Liu}}, \bibinfo {author} {\bibfnamefont {H.}~\bibnamefont {Lei}},  \emph {et~al.},\ }\href@noop {} {\bibfield  {journal} {\bibinfo  {journal} {Nature communications}\ }\textbf {\bibinfo {volume} {12}},\ \bibinfo {pages} {3129} (\bibinfo {year} {2021})}\BibitemShut {NoStop}%
\bibitem [{\citenamefont {Nwaogbo}\ \emph {et~al.}(2025{\natexlab{b}})\citenamefont {Nwaogbo}, \citenamefont {Das},\ and\ \citenamefont {Ekuma}}]{NWAOGBO2025101780}%
  \BibitemOpen
  \bibfield  {author} {\bibinfo {author} {\bibfnamefont {C.~I.}\ \bibnamefont {Nwaogbo}}, \bibinfo {author} {\bibfnamefont {S.~K.}\ \bibnamefont {Das}}, \ and\ \bibinfo {author} {\bibfnamefont {C.~E.}\ \bibnamefont {Ekuma}},\ }\href {\doibase https://doi.org/10.1016/j.mtphys.2025.101780} {\bibfield  {journal} {\bibinfo  {journal} {Materials Today Physics}\ }\textbf {\bibinfo {volume} {57}},\ \bibinfo {pages} {101780} (\bibinfo {year} {2025}{\natexlab{b}})}\BibitemShut {NoStop}%
\bibitem [{\citenamefont {Zhu}\ \emph {et~al.}(2025)\citenamefont {Zhu}, \citenamefont {Yuan},\ and\ \citenamefont {Wang}}]{zhu2025theoretical}%
  \BibitemOpen
  \bibfield  {author} {\bibinfo {author} {\bibfnamefont {Y.}~\bibnamefont {Zhu}}, \bibinfo {author} {\bibfnamefont {J.-H.}\ \bibnamefont {Yuan}}, \ and\ \bibinfo {author} {\bibfnamefont {J.}~\bibnamefont {Wang}},\ }\href@noop {} {\bibfield  {journal} {\bibinfo  {journal} {Applied Surface Science}\ ,\ \bibinfo {pages} {163829}} (\bibinfo {year} {2025})}\BibitemShut {NoStop}%
\bibitem [{\citenamefont {Zhu}\ \emph {et~al.}(2023)\citenamefont {Zhu}, \citenamefont {Yuan}, \citenamefont {Fang}, \citenamefont {Sun},\ and\ \citenamefont {Wang}}]{zhu2023family}%
  \BibitemOpen
  \bibfield  {author} {\bibinfo {author} {\bibfnamefont {Y.}~\bibnamefont {Zhu}}, \bibinfo {author} {\bibfnamefont {J.-H.}\ \bibnamefont {Yuan}}, \bibinfo {author} {\bibfnamefont {W.-Y.}\ \bibnamefont {Fang}}, \bibinfo {author} {\bibfnamefont {Z.-G.}\ \bibnamefont {Sun}}, \ and\ \bibinfo {author} {\bibfnamefont {J.}~\bibnamefont {Wang}},\ }\href@noop {} {\bibfield  {journal} {\bibinfo  {journal} {Applied Surface Science}\ }\textbf {\bibinfo {volume} {636}},\ \bibinfo {pages} {157817} (\bibinfo {year} {2023})}\BibitemShut {NoStop}%
\bibitem [{\citenamefont {Ohta}\ \emph {et~al.}(2006)\citenamefont {Ohta}, \citenamefont {Bostwick}, \citenamefont {Seyller}, \citenamefont {Horn},\ and\ \citenamefont {Rotenberg}}]{ohta2006controlling}%
  \BibitemOpen
  \bibfield  {author} {\bibinfo {author} {\bibfnamefont {T.}~\bibnamefont {Ohta}}, \bibinfo {author} {\bibfnamefont {A.}~\bibnamefont {Bostwick}}, \bibinfo {author} {\bibfnamefont {T.}~\bibnamefont {Seyller}}, \bibinfo {author} {\bibfnamefont {K.}~\bibnamefont {Horn}}, \ and\ \bibinfo {author} {\bibfnamefont {E.}~\bibnamefont {Rotenberg}},\ }\href@noop {} {\bibfield  {journal} {\bibinfo  {journal} {Science}\ }\textbf {\bibinfo {volume} {313}},\ \bibinfo {pages} {951} (\bibinfo {year} {2006})}\BibitemShut {NoStop}%
\bibitem [{\citenamefont {Bernevig}\ and\ \citenamefont {Efetov}(2024)}]{bernevig2024twisted}%
  \BibitemOpen
  \bibfield  {author} {\bibinfo {author} {\bibfnamefont {B.~A.}\ \bibnamefont {Bernevig}}\ and\ \bibinfo {author} {\bibfnamefont {D.~K.}\ \bibnamefont {Efetov}},\ }\href@noop {} {\bibfield  {journal} {\bibinfo  {journal} {Physics Today}\ }\textbf {\bibinfo {volume} {77}},\ \bibinfo {pages} {38} (\bibinfo {year} {2024})}\BibitemShut {NoStop}%
\bibitem [{\citenamefont {Su}\ \emph {et~al.}(2016)\citenamefont {Su}, \citenamefont {Ju}, \citenamefont {Zhang}, \citenamefont {Guo}, \citenamefont {Yong}, \citenamefont {Cui},\ and\ \citenamefont {Li}}]{su2016band}%
  \BibitemOpen
  \bibfield  {author} {\bibinfo {author} {\bibfnamefont {X.}~\bibnamefont {Su}}, \bibinfo {author} {\bibfnamefont {W.}~\bibnamefont {Ju}}, \bibinfo {author} {\bibfnamefont {R.}~\bibnamefont {Zhang}}, \bibinfo {author} {\bibfnamefont {C.}~\bibnamefont {Guo}}, \bibinfo {author} {\bibfnamefont {Y.}~\bibnamefont {Yong}}, \bibinfo {author} {\bibfnamefont {H.}~\bibnamefont {Cui}}, \ and\ \bibinfo {author} {\bibfnamefont {X.}~\bibnamefont {Li}},\ }\href@noop {} {\bibfield  {journal} {\bibinfo  {journal} {Physica E: Low-dimensional Systems and Nanostructures}\ }\textbf {\bibinfo {volume} {84}},\ \bibinfo {pages} {216} (\bibinfo {year} {2016})}\BibitemShut {NoStop}%
\bibitem [{\citenamefont {Peng}\ \emph {et~al.}(2020)\citenamefont {Peng}, \citenamefont {Chen}, \citenamefont {Fan}, \citenamefont {Srolovitz},\ and\ \citenamefont {Lei}}]{peng2020strain}%
  \BibitemOpen
  \bibfield  {author} {\bibinfo {author} {\bibfnamefont {Z.}~\bibnamefont {Peng}}, \bibinfo {author} {\bibfnamefont {X.}~\bibnamefont {Chen}}, \bibinfo {author} {\bibfnamefont {Y.}~\bibnamefont {Fan}}, \bibinfo {author} {\bibfnamefont {D.~J.}\ \bibnamefont {Srolovitz}}, \ and\ \bibinfo {author} {\bibfnamefont {D.}~\bibnamefont {Lei}},\ }\href@noop {} {\bibfield  {journal} {\bibinfo  {journal} {Light: Science \& Applications}\ }\textbf {\bibinfo {volume} {9}},\ \bibinfo {pages} {190} (\bibinfo {year} {2020})}\BibitemShut {NoStop}%
\bibitem [{\citenamefont {Maesato}\ \emph {et~al.}(2000)\citenamefont {Maesato}, \citenamefont {Kaga}, \citenamefont {Kondo},\ and\ \citenamefont {Kagoshima}}]{maesato2000uniaxial}%
  \BibitemOpen
  \bibfield  {author} {\bibinfo {author} {\bibfnamefont {M.}~\bibnamefont {Maesato}}, \bibinfo {author} {\bibfnamefont {Y.}~\bibnamefont {Kaga}}, \bibinfo {author} {\bibfnamefont {R.}~\bibnamefont {Kondo}}, \ and\ \bibinfo {author} {\bibfnamefont {S.}~\bibnamefont {Kagoshima}},\ }\href@noop {} {\bibfield  {journal} {\bibinfo  {journal} {Review of Scientific Instruments}\ }\textbf {\bibinfo {volume} {71}},\ \bibinfo {pages} {176} (\bibinfo {year} {2000})}\BibitemShut {NoStop}%
\bibitem [{\citenamefont {Yu}\ \emph {et~al.}(2024)\citenamefont {Yu}, \citenamefont {Peng}, \citenamefont {Xu}, \citenamefont {Shi}, \citenamefont {Li}, \citenamefont {Meng}, \citenamefont {He}, \citenamefont {Wang}, \citenamefont {Duan}, \citenamefont {Tong} \emph {et~al.}}]{yu2024manipulating}%
  \BibitemOpen
  \bibfield  {author} {\bibinfo {author} {\bibfnamefont {X.}~\bibnamefont {Yu}}, \bibinfo {author} {\bibfnamefont {Z.}~\bibnamefont {Peng}}, \bibinfo {author} {\bibfnamefont {L.}~\bibnamefont {Xu}}, \bibinfo {author} {\bibfnamefont {W.}~\bibnamefont {Shi}}, \bibinfo {author} {\bibfnamefont {Z.}~\bibnamefont {Li}}, \bibinfo {author} {\bibfnamefont {X.}~\bibnamefont {Meng}}, \bibinfo {author} {\bibfnamefont {X.}~\bibnamefont {He}}, \bibinfo {author} {\bibfnamefont {Z.}~\bibnamefont {Wang}}, \bibinfo {author} {\bibfnamefont {S.}~\bibnamefont {Duan}}, \bibinfo {author} {\bibfnamefont {L.}~\bibnamefont {Tong}},  \emph {et~al.},\ }\href@noop {} {\bibfield  {journal} {\bibinfo  {journal} {Small}\ }\textbf {\bibinfo {volume} {20}},\ \bibinfo {pages} {2402561} (\bibinfo {year} {2024})}\BibitemShut {NoStop}%
\bibitem [{\citenamefont {Azizimanesh}\ \emph {et~al.}(2021)\citenamefont {Azizimanesh}, \citenamefont {Pe{\~n}a}, \citenamefont {Sewaket}, \citenamefont {Hou},\ and\ \citenamefont {Wu}}]{azizimanesh2021uniaxial}%
  \BibitemOpen
  \bibfield  {author} {\bibinfo {author} {\bibfnamefont {A.}~\bibnamefont {Azizimanesh}}, \bibinfo {author} {\bibfnamefont {T.}~\bibnamefont {Pe{\~n}a}}, \bibinfo {author} {\bibfnamefont {A.}~\bibnamefont {Sewaket}}, \bibinfo {author} {\bibfnamefont {W.}~\bibnamefont {Hou}}, \ and\ \bibinfo {author} {\bibfnamefont {S.~M.}\ \bibnamefont {Wu}},\ }\href@noop {} {\bibfield  {journal} {\bibinfo  {journal} {Applied Physics Letters}\ }\textbf {\bibinfo {volume} {118}} (\bibinfo {year} {2021})}\BibitemShut {NoStop}%
\bibitem [{\citenamefont {Kresse}\ and\ \citenamefont {Furthm{\"u}ller}(1996{\natexlab{a}})}]{kresse1996efficiency}%
  \BibitemOpen
  \bibfield  {author} {\bibinfo {author} {\bibfnamefont {G.}~\bibnamefont {Kresse}}\ and\ \bibinfo {author} {\bibfnamefont {J.}~\bibnamefont {Furthm{\"u}ller}},\ }\href@noop {} {\bibfield  {journal} {\bibinfo  {journal} {Computational materials science}\ }\textbf {\bibinfo {volume} {6}},\ \bibinfo {pages} {15} (\bibinfo {year} {1996}{\natexlab{a}})}\BibitemShut {NoStop}%
\bibitem [{\citenamefont {Kresse}\ and\ \citenamefont {Furthm{\"u}ller}(1996{\natexlab{b}})}]{kresse1996efficient}%
  \BibitemOpen
  \bibfield  {author} {\bibinfo {author} {\bibfnamefont {G.}~\bibnamefont {Kresse}}\ and\ \bibinfo {author} {\bibfnamefont {J.}~\bibnamefont {Furthm{\"u}ller}},\ }\href@noop {} {\bibfield  {journal} {\bibinfo  {journal} {Physical review B}\ }\textbf {\bibinfo {volume} {54}},\ \bibinfo {pages} {11169} (\bibinfo {year} {1996}{\natexlab{b}})}\BibitemShut {NoStop}%
\bibitem [{\citenamefont {Perdew}\ \emph {et~al.}(1996)\citenamefont {Perdew}, \citenamefont {Burke},\ and\ \citenamefont {Ernzerhof}}]{perdew1996generalized}%
  \BibitemOpen
  \bibfield  {author} {\bibinfo {author} {\bibfnamefont {J.~P.}\ \bibnamefont {Perdew}}, \bibinfo {author} {\bibfnamefont {K.}~\bibnamefont {Burke}}, \ and\ \bibinfo {author} {\bibfnamefont {M.}~\bibnamefont {Ernzerhof}},\ }\href@noop {} {\bibfield  {journal} {\bibinfo  {journal} {Physical review letters}\ }\textbf {\bibinfo {volume} {77}},\ \bibinfo {pages} {3865} (\bibinfo {year} {1996})}\BibitemShut {NoStop}%
\bibitem [{\citenamefont {Liu}\ \emph {et~al.}(2022)\citenamefont {Liu}, \citenamefont {Ekuma}, \citenamefont {Li}, \citenamefont {Yang},\ and\ \citenamefont {Li}}]{liu2022elastool}%
  \BibitemOpen
  \bibfield  {author} {\bibinfo {author} {\bibfnamefont {Z.-L.}\ \bibnamefont {Liu}}, \bibinfo {author} {\bibfnamefont {C.}~\bibnamefont {Ekuma}}, \bibinfo {author} {\bibfnamefont {W.-Q.}\ \bibnamefont {Li}}, \bibinfo {author} {\bibfnamefont {J.-Q.}\ \bibnamefont {Yang}}, \ and\ \bibinfo {author} {\bibfnamefont {X.-J.}\ \bibnamefont {Li}},\ }\href@noop {} {\bibfield  {journal} {\bibinfo  {journal} {Computer Physics Communications}\ }\textbf {\bibinfo {volume} {270}},\ \bibinfo {pages} {108180} (\bibinfo {year} {2022})}\BibitemShut {NoStop}%
\bibitem [{\citenamefont {Ekuma}\ and\ \citenamefont {Liu}(2024)}]{EKUMA2024109161}%
  \BibitemOpen
  \bibfield  {author} {\bibinfo {author} {\bibfnamefont {C.}~\bibnamefont {Ekuma}}\ and\ \bibinfo {author} {\bibfnamefont {Z.-L.}\ \bibnamefont {Liu}},\ }\href {\doibase https://doi.org/10.1016/j.cpc.2024.109161} {\bibfield  {journal} {\bibinfo  {journal} {Computer Physics Communications}\ }\textbf {\bibinfo {volume} {300}},\ \bibinfo {pages} {109161} (\bibinfo {year} {2024})}\BibitemShut {NoStop}%
\bibitem [{\citenamefont {Lalmi}\ \emph {et~al.}(2010)\citenamefont {Lalmi}, \citenamefont {Oughaddou}, \citenamefont {Enriquez}, \citenamefont {Kara}, \citenamefont {Vizzini}, \citenamefont {Ealet},\ and\ \citenamefont {Aufray}}]{lalmi2010epitaxial}%
  \BibitemOpen
  \bibfield  {author} {\bibinfo {author} {\bibfnamefont {B.}~\bibnamefont {Lalmi}}, \bibinfo {author} {\bibfnamefont {H.}~\bibnamefont {Oughaddou}}, \bibinfo {author} {\bibfnamefont {H.}~\bibnamefont {Enriquez}}, \bibinfo {author} {\bibfnamefont {A.}~\bibnamefont {Kara}}, \bibinfo {author} {\bibfnamefont {S.}~\bibnamefont {Vizzini}}, \bibinfo {author} {\bibfnamefont {B.}~\bibnamefont {Ealet}}, \ and\ \bibinfo {author} {\bibfnamefont {B.}~\bibnamefont {Aufray}},\ }\href@noop {} {\bibfield  {journal} {\bibinfo  {journal} {Applied Physics Letters}\ }\textbf {\bibinfo {volume} {97}} (\bibinfo {year} {2010})}\BibitemShut {NoStop}%
\bibitem [{\citenamefont {Manni}\ \emph {et~al.}(2017)\citenamefont {Manni}, \citenamefont {Bud'ko},\ and\ \citenamefont {Canfield}}]{manni2017gdptpb}%
  \BibitemOpen
  \bibfield  {author} {\bibinfo {author} {\bibfnamefont {S.}~\bibnamefont {Manni}}, \bibinfo {author} {\bibfnamefont {S.~L.}\ \bibnamefont {Bud'ko}}, \ and\ \bibinfo {author} {\bibfnamefont {P.~C.}\ \bibnamefont {Canfield}},\ }\href@noop {} {\bibfield  {journal} {\bibinfo  {journal} {Physical Review B}\ }\textbf {\bibinfo {volume} {96}},\ \bibinfo {pages} {054435} (\bibinfo {year} {2017})}\BibitemShut {NoStop}%
\bibitem [{\citenamefont {Lee}\ \emph {et~al.}(2012)\citenamefont {Lee}, \citenamefont {Zhang}, \citenamefont {Zhang}, \citenamefont {Chang}, \citenamefont {Lin}, \citenamefont {Chang}, \citenamefont {Yu}, \citenamefont {Wang}, \citenamefont {Chang}, \citenamefont {Li} \emph {et~al.}}]{lee2012synthesis}%
  \BibitemOpen
  \bibfield  {author} {\bibinfo {author} {\bibfnamefont {Y.-H.}\ \bibnamefont {Lee}}, \bibinfo {author} {\bibfnamefont {X.-Q.}\ \bibnamefont {Zhang}}, \bibinfo {author} {\bibfnamefont {W.}~\bibnamefont {Zhang}}, \bibinfo {author} {\bibfnamefont {M.-T.}\ \bibnamefont {Chang}}, \bibinfo {author} {\bibfnamefont {C.-T.}\ \bibnamefont {Lin}}, \bibinfo {author} {\bibfnamefont {K.-D.}\ \bibnamefont {Chang}}, \bibinfo {author} {\bibfnamefont {Y.-C.}\ \bibnamefont {Yu}}, \bibinfo {author} {\bibfnamefont {J.~T.-W.}\ \bibnamefont {Wang}}, \bibinfo {author} {\bibfnamefont {C.-S.}\ \bibnamefont {Chang}}, \bibinfo {author} {\bibfnamefont {L.-J.}\ \bibnamefont {Li}},  \emph {et~al.},\ }\href@noop {} {\bibfield  {journal} {\bibinfo  {journal} {arXiv preprint arXiv:1202.5458}\ } (\bibinfo {year} {2012})}\BibitemShut {NoStop}%
\bibitem [{\citenamefont {Born}\ and\ \citenamefont {Huang}(1996)}]{born1996dynamical}%
  \BibitemOpen
  \bibfield  {author} {\bibinfo {author} {\bibfnamefont {M.}~\bibnamefont {Born}}\ and\ \bibinfo {author} {\bibfnamefont {K.}~\bibnamefont {Huang}},\ }\href@noop {} {\emph {\bibinfo {title} {Dynamical theory of crystal lattices}}}\ (\bibinfo  {publisher} {Oxford university press},\ \bibinfo {year} {1996})\BibitemShut {NoStop}%
\bibitem [{\citenamefont {Mouhat}\ and\ \citenamefont {Coudert}(2014)}]{PhysRevB.90.224104}%
  \BibitemOpen
  \bibfield  {author} {\bibinfo {author} {\bibfnamefont {F.}~\bibnamefont {Mouhat}}\ and\ \bibinfo {author} {\bibfnamefont {F.~m. c.-X.}\ \bibnamefont {Coudert}},\ }\href {\doibase 10.1103/PhysRevB.90.224104} {\bibfield  {journal} {\bibinfo  {journal} {Phys. Rev. B}\ }\textbf {\bibinfo {volume} {90}},\ \bibinfo {pages} {224104} (\bibinfo {year} {2014})}\BibitemShut {NoStop}%
\bibitem [{\citenamefont {Cadelano}\ \emph {et~al.}(2010)\citenamefont {Cadelano}, \citenamefont {Palla}, \citenamefont {Giordano},\ and\ \citenamefont {Colombo}}]{PhysRevB.82.235414}%
  \BibitemOpen
  \bibfield  {author} {\bibinfo {author} {\bibfnamefont {E.}~\bibnamefont {Cadelano}}, \bibinfo {author} {\bibfnamefont {P.~L.}\ \bibnamefont {Palla}}, \bibinfo {author} {\bibfnamefont {S.}~\bibnamefont {Giordano}}, \ and\ \bibinfo {author} {\bibfnamefont {L.}~\bibnamefont {Colombo}},\ }\href {\doibase 10.1103/PhysRevB.82.235414} {\bibfield  {journal} {\bibinfo  {journal} {Phys. Rev. B}\ }\textbf {\bibinfo {volume} {82}},\ \bibinfo {pages} {235414} (\bibinfo {year} {2010})}\BibitemShut {NoStop}%
\bibitem [{\citenamefont {Andrew}\ \emph {et~al.}(2012)\citenamefont {Andrew}, \citenamefont {Mapasha}, \citenamefont {Ukpong},\ and\ \citenamefont {Chetty}}]{andrew2012mechanical}%
  \BibitemOpen
  \bibfield  {author} {\bibinfo {author} {\bibfnamefont {R.~C.}\ \bibnamefont {Andrew}}, \bibinfo {author} {\bibfnamefont {R.~E.}\ \bibnamefont {Mapasha}}, \bibinfo {author} {\bibfnamefont {A.~M.}\ \bibnamefont {Ukpong}}, \ and\ \bibinfo {author} {\bibfnamefont {N.}~\bibnamefont {Chetty}},\ }\href@noop {} {\bibfield  {journal} {\bibinfo  {journal} {Physical Review B—Condensed Matter and Materials Physics}\ }\textbf {\bibinfo {volume} {85}},\ \bibinfo {pages} {125428} (\bibinfo {year} {2012})}\BibitemShut {NoStop}%
\bibitem [{\citenamefont {Cooper}\ \emph {et~al.}(2013)\citenamefont {Cooper}, \citenamefont {Lee}, \citenamefont {Marianetti}, \citenamefont {Wei}, \citenamefont {Hone},\ and\ \citenamefont {Kysar}}]{cooper2013nonlinear}%
  \BibitemOpen
  \bibfield  {author} {\bibinfo {author} {\bibfnamefont {R.~C.}\ \bibnamefont {Cooper}}, \bibinfo {author} {\bibfnamefont {C.}~\bibnamefont {Lee}}, \bibinfo {author} {\bibfnamefont {C.~A.}\ \bibnamefont {Marianetti}}, \bibinfo {author} {\bibfnamefont {X.}~\bibnamefont {Wei}}, \bibinfo {author} {\bibfnamefont {J.}~\bibnamefont {Hone}}, \ and\ \bibinfo {author} {\bibfnamefont {J.~W.}\ \bibnamefont {Kysar}},\ }\href@noop {} {\bibfield  {journal} {\bibinfo  {journal} {Physical Review B—Condensed Matter and Materials Physics}\ }\textbf {\bibinfo {volume} {87}},\ \bibinfo {pages} {035423} (\bibinfo {year} {2013})}\BibitemShut {NoStop}%
\bibitem [{\citenamefont {Castellanos-Gomez}\ \emph {et~al.}(2012)\citenamefont {Castellanos-Gomez}, \citenamefont {Poot}, \citenamefont {Steele}, \citenamefont {Van Der~Zant}, \citenamefont {Agra{\"\i}t},\ and\ \citenamefont {Rubio-Bollinger}}]{castellanos2012elastic}%
  \BibitemOpen
  \bibfield  {author} {\bibinfo {author} {\bibfnamefont {A.}~\bibnamefont {Castellanos-Gomez}}, \bibinfo {author} {\bibfnamefont {M.}~\bibnamefont {Poot}}, \bibinfo {author} {\bibfnamefont {G.~A.}\ \bibnamefont {Steele}}, \bibinfo {author} {\bibfnamefont {H.~S.}\ \bibnamefont {Van Der~Zant}}, \bibinfo {author} {\bibfnamefont {N.}~\bibnamefont {Agra{\"\i}t}}, \ and\ \bibinfo {author} {\bibfnamefont {G.}~\bibnamefont {Rubio-Bollinger}},\ }\href@noop {} {\bibfield  {journal} {\bibinfo  {journal} {arXiv preprint arXiv:1202.4439}\ } (\bibinfo {year} {2012})}\BibitemShut {NoStop}%
\bibitem [{\citenamefont {Bertolazzi}\ \emph {et~al.}(2011)\citenamefont {Bertolazzi}, \citenamefont {Brivio},\ and\ \citenamefont {Kis}}]{bertolazzi2011stretching}%
  \BibitemOpen
  \bibfield  {author} {\bibinfo {author} {\bibfnamefont {S.}~\bibnamefont {Bertolazzi}}, \bibinfo {author} {\bibfnamefont {J.}~\bibnamefont {Brivio}}, \ and\ \bibinfo {author} {\bibfnamefont {A.}~\bibnamefont {Kis}},\ }\href@noop {} {\bibfield  {journal} {\bibinfo  {journal} {ACS nano}\ }\textbf {\bibinfo {volume} {5}},\ \bibinfo {pages} {9703} (\bibinfo {year} {2011})}\BibitemShut {NoStop}%
\bibitem [{\citenamefont {Jiang}\ and\ \citenamefont {Park}(2014)}]{jiang2014mechanical}%
  \BibitemOpen
  \bibfield  {author} {\bibinfo {author} {\bibfnamefont {J.-W.}\ \bibnamefont {Jiang}}\ and\ \bibinfo {author} {\bibfnamefont {H.~S.}\ \bibnamefont {Park}},\ }\href@noop {} {\bibfield  {journal} {\bibinfo  {journal} {Journal of Physics D: Applied Physics}\ }\textbf {\bibinfo {volume} {47}},\ \bibinfo {pages} {385304} (\bibinfo {year} {2014})}\BibitemShut {NoStop}%
\bibitem [{\citenamefont {Oriola}\ \emph {et~al.}(2025)\citenamefont {Oriola}, \citenamefont {Maile}, \citenamefont {Nguyen},\ and\ \citenamefont {Payton}}]{oriola2025toward}%
  \BibitemOpen
  \bibfield  {author} {\bibinfo {author} {\bibfnamefont {A.}~\bibnamefont {Oriola}}, \bibinfo {author} {\bibfnamefont {J.}~\bibnamefont {Maile}}, \bibinfo {author} {\bibfnamefont {A.}~\bibnamefont {Nguyen}}, \ and\ \bibinfo {author} {\bibfnamefont {E.}~\bibnamefont {Payton}},\ }\href@noop {} {\bibfield  {journal} {\bibinfo  {journal} {JOM}\ ,\ \bibinfo {pages} {1}} (\bibinfo {year} {2025})}\BibitemShut {NoStop}%
\bibitem [{\citenamefont {Pugh}(1954)}]{pugh1954xcii}%
  \BibitemOpen
  \bibfield  {author} {\bibinfo {author} {\bibfnamefont {S.}~\bibnamefont {Pugh}},\ }\href@noop {} {\bibfield  {journal} {\bibinfo  {journal} {The London, Edinburgh, and Dublin Philosophical Magazine and Journal of Science}\ }\textbf {\bibinfo {volume} {45}},\ \bibinfo {pages} {823} (\bibinfo {year} {1954})}\BibitemShut {NoStop}%
\bibitem [{\citenamefont {Frantsevich}(1982)}]{frantsevich1982elastic}%
  \BibitemOpen
  \bibfield  {author} {\bibinfo {author} {\bibfnamefont {I.~N.}\ \bibnamefont {Frantsevich}},\ }\href@noop {} {\bibfield  {journal} {\bibinfo  {journal} {Reference book}\ } (\bibinfo {year} {1982})}\BibitemShut {NoStop}%
\bibitem [{\citenamefont {Zhou}\ \emph {et~al.}(2014)\citenamefont {Zhou}, \citenamefont {Liu}, \citenamefont {Ming}, \citenamefont {Wang},\ and\ \citenamefont {Liu}}]{zhou2014sd}%
  \BibitemOpen
  \bibfield  {author} {\bibinfo {author} {\bibfnamefont {M.}~\bibnamefont {Zhou}}, \bibinfo {author} {\bibfnamefont {Z.}~\bibnamefont {Liu}}, \bibinfo {author} {\bibfnamefont {W.}~\bibnamefont {Ming}}, \bibinfo {author} {\bibfnamefont {Z.}~\bibnamefont {Wang}}, \ and\ \bibinfo {author} {\bibfnamefont {F.}~\bibnamefont {Liu}},\ }\href@noop {} {\bibfield  {journal} {\bibinfo  {journal} {Physical review letters}\ }\textbf {\bibinfo {volume} {113}},\ \bibinfo {pages} {236802} (\bibinfo {year} {2014})}\BibitemShut {NoStop}%
\bibitem [{\citenamefont {Yin}\ \emph {et~al.}(2019)\citenamefont {Yin}, \citenamefont {Zhang}, \citenamefont {Chang}, \citenamefont {Wang}, \citenamefont {Tsirkin}, \citenamefont {Guguchia}, \citenamefont {Lian}, \citenamefont {Zhou}, \citenamefont {Jiang}, \citenamefont {Belopolski} \emph {et~al.}}]{yin2019negative}%
  \BibitemOpen
  \bibfield  {author} {\bibinfo {author} {\bibfnamefont {J.-X.}\ \bibnamefont {Yin}}, \bibinfo {author} {\bibfnamefont {S.~S.}\ \bibnamefont {Zhang}}, \bibinfo {author} {\bibfnamefont {G.}~\bibnamefont {Chang}}, \bibinfo {author} {\bibfnamefont {Q.}~\bibnamefont {Wang}}, \bibinfo {author} {\bibfnamefont {S.~S.}\ \bibnamefont {Tsirkin}}, \bibinfo {author} {\bibfnamefont {Z.}~\bibnamefont {Guguchia}}, \bibinfo {author} {\bibfnamefont {B.}~\bibnamefont {Lian}}, \bibinfo {author} {\bibfnamefont {H.}~\bibnamefont {Zhou}}, \bibinfo {author} {\bibfnamefont {K.}~\bibnamefont {Jiang}}, \bibinfo {author} {\bibfnamefont {I.}~\bibnamefont {Belopolski}},  \emph {et~al.},\ }\href@noop {} {\bibfield  {journal} {\bibinfo  {journal} {Nature Physics}\ }\textbf {\bibinfo {volume} {15}},\ \bibinfo {pages} {443} (\bibinfo {year} {2019})}\BibitemShut {NoStop}%
\bibitem [{\citenamefont {Park}\ \emph {et~al.}(2020)\citenamefont {Park}, \citenamefont {Kang}, \citenamefont {Kim}, \citenamefont {Lee}, \citenamefont {Kim}, \citenamefont {Sim}, \citenamefont {Lee}, \citenamefont {Karuppannan}, \citenamefont {Kim}, \citenamefont {Kim} \emph {et~al.}}]{park2020kagome}%
  \BibitemOpen
  \bibfield  {author} {\bibinfo {author} {\bibfnamefont {S.}~\bibnamefont {Park}}, \bibinfo {author} {\bibfnamefont {S.}~\bibnamefont {Kang}}, \bibinfo {author} {\bibfnamefont {H.}~\bibnamefont {Kim}}, \bibinfo {author} {\bibfnamefont {K.~H.}\ \bibnamefont {Lee}}, \bibinfo {author} {\bibfnamefont {P.}~\bibnamefont {Kim}}, \bibinfo {author} {\bibfnamefont {S.}~\bibnamefont {Sim}}, \bibinfo {author} {\bibfnamefont {N.}~\bibnamefont {Lee}}, \bibinfo {author} {\bibfnamefont {B.}~\bibnamefont {Karuppannan}}, \bibinfo {author} {\bibfnamefont {J.}~\bibnamefont {Kim}}, \bibinfo {author} {\bibfnamefont {J.}~\bibnamefont {Kim}},  \emph {et~al.},\ }\href@noop {} {\bibfield  {journal} {\bibinfo  {journal} {Scientific reports}\ }\textbf {\bibinfo {volume} {10}},\ \bibinfo {pages} {20998} (\bibinfo {year} {2020})}\BibitemShut {NoStop}%
\bibitem [{\citenamefont {Li}\ \emph {et~al.}(2025)\citenamefont {Li}, \citenamefont {Ma}, \citenamefont {Lou},\ and\ \citenamefont {Wang}}]{li2025electronic}%
  \BibitemOpen
  \bibfield  {author} {\bibinfo {author} {\bibfnamefont {M.}~\bibnamefont {Li}}, \bibinfo {author} {\bibfnamefont {H.}~\bibnamefont {Ma}}, \bibinfo {author} {\bibfnamefont {R.}~\bibnamefont {Lou}}, \ and\ \bibinfo {author} {\bibfnamefont {S.}~\bibnamefont {Wang}},\ }\href@noop {} {\bibfield  {journal} {\bibinfo  {journal} {Chinese Physics B}\ }\textbf {\bibinfo {volume} {34}},\ \bibinfo {pages} {017101} (\bibinfo {year} {2025})}\BibitemShut {NoStop}%
\bibitem [{\citenamefont {Cusati}\ \emph {et~al.}(2018)\citenamefont {Cusati}, \citenamefont {Fortunelli}, \citenamefont {Fiori},\ and\ \citenamefont {Iannaccone}}]{cusati2018stacking}%
  \BibitemOpen
  \bibfield  {author} {\bibinfo {author} {\bibfnamefont {T.}~\bibnamefont {Cusati}}, \bibinfo {author} {\bibfnamefont {A.}~\bibnamefont {Fortunelli}}, \bibinfo {author} {\bibfnamefont {G.}~\bibnamefont {Fiori}}, \ and\ \bibinfo {author} {\bibfnamefont {G.}~\bibnamefont {Iannaccone}},\ }\href@noop {} {\bibfield  {journal} {\bibinfo  {journal} {Physical Review B}\ }\textbf {\bibinfo {volume} {98}},\ \bibinfo {pages} {115403} (\bibinfo {year} {2018})}\BibitemShut {NoStop}%
\bibitem [{\citenamefont {Ma}\ \emph {et~al.}(2014)\citenamefont {Ma}, \citenamefont {Hu}, \citenamefont {Zhao}, \citenamefont {Tang}, \citenamefont {Wu}, \citenamefont {Zhou},\ and\ \citenamefont {Zhang}}]{ma2014tunable}%
  \BibitemOpen
  \bibfield  {author} {\bibinfo {author} {\bibfnamefont {Z.}~\bibnamefont {Ma}}, \bibinfo {author} {\bibfnamefont {Z.}~\bibnamefont {Hu}}, \bibinfo {author} {\bibfnamefont {X.}~\bibnamefont {Zhao}}, \bibinfo {author} {\bibfnamefont {Q.}~\bibnamefont {Tang}}, \bibinfo {author} {\bibfnamefont {D.}~\bibnamefont {Wu}}, \bibinfo {author} {\bibfnamefont {Z.}~\bibnamefont {Zhou}}, \ and\ \bibinfo {author} {\bibfnamefont {L.}~\bibnamefont {Zhang}},\ }\href@noop {} {\bibfield  {journal} {\bibinfo  {journal} {The Journal of Physical Chemistry C}\ }\textbf {\bibinfo {volume} {118}},\ \bibinfo {pages} {5593} (\bibinfo {year} {2014})}\BibitemShut {NoStop}%
\bibitem [{\citenamefont {Hu}\ \emph {et~al.}(2025)\citenamefont {Hu}, \citenamefont {Xiong}, \citenamefont {Zhao},\ and\ \citenamefont {Chen}}]{hu2025tunable}%
  \BibitemOpen
  \bibfield  {author} {\bibinfo {author} {\bibfnamefont {H.}~\bibnamefont {Hu}}, \bibinfo {author} {\bibfnamefont {Z.}~\bibnamefont {Xiong}}, \bibinfo {author} {\bibfnamefont {J.}~\bibnamefont {Zhao}}, \ and\ \bibinfo {author} {\bibfnamefont {L.}~\bibnamefont {Chen}},\ }\href@noop {} {\bibfield  {journal} {\bibinfo  {journal} {Surface Science}\ ,\ \bibinfo {pages} {122833}} (\bibinfo {year} {2025})}\BibitemShut {NoStop}%
\bibitem [{\citenamefont {Yun}\ \emph {et~al.}(2018)\citenamefont {Yun}, \citenamefont {Zhang}, \citenamefont {Ren}, \citenamefont {Xu}, \citenamefont {Yan}, \citenamefont {Zhao},\ and\ \citenamefont {Zhang}}]{yun2018tunable}%
  \BibitemOpen
  \bibfield  {author} {\bibinfo {author} {\bibfnamefont {J.}~\bibnamefont {Yun}}, \bibinfo {author} {\bibfnamefont {Y.}~\bibnamefont {Zhang}}, \bibinfo {author} {\bibfnamefont {Y.}~\bibnamefont {Ren}}, \bibinfo {author} {\bibfnamefont {M.}~\bibnamefont {Xu}}, \bibinfo {author} {\bibfnamefont {J.}~\bibnamefont {Yan}}, \bibinfo {author} {\bibfnamefont {W.}~\bibnamefont {Zhao}}, \ and\ \bibinfo {author} {\bibfnamefont {Z.}~\bibnamefont {Zhang}},\ }\href@noop {} {\bibfield  {journal} {\bibinfo  {journal} {Physical Chemistry Chemical Physics}\ }\textbf {\bibinfo {volume} {20}},\ \bibinfo {pages} {26934} (\bibinfo {year} {2018})}\BibitemShut {NoStop}%
\bibitem [{\citenamefont {Zou}\ \emph {et~al.}(2011)\citenamefont {Zou}, \citenamefont {Hong},\ and\ \citenamefont {Zhu}}]{zou2011effective}%
  \BibitemOpen
  \bibfield  {author} {\bibinfo {author} {\bibfnamefont {K.}~\bibnamefont {Zou}}, \bibinfo {author} {\bibfnamefont {X.}~\bibnamefont {Hong}}, \ and\ \bibinfo {author} {\bibfnamefont {J.}~\bibnamefont {Zhu}},\ }\href@noop {} {\bibfield  {journal} {\bibinfo  {journal} {Physical Review B—Condensed Matter and Materials Physics}\ }\textbf {\bibinfo {volume} {84}},\ \bibinfo {pages} {085408} (\bibinfo {year} {2011})}\BibitemShut {NoStop}%
\bibitem [{\citenamefont {He}\ \emph {et~al.}(2014)\citenamefont {He}, \citenamefont {Hummer},\ and\ \citenamefont {Franchini}}]{he2014stacking}%
  \BibitemOpen
  \bibfield  {author} {\bibinfo {author} {\bibfnamefont {J.}~\bibnamefont {He}}, \bibinfo {author} {\bibfnamefont {K.}~\bibnamefont {Hummer}}, \ and\ \bibinfo {author} {\bibfnamefont {C.}~\bibnamefont {Franchini}},\ }\href@noop {} {\bibfield  {journal} {\bibinfo  {journal} {Physical Review B}\ }\textbf {\bibinfo {volume} {89}},\ \bibinfo {pages} {075409} (\bibinfo {year} {2014})}\BibitemShut {NoStop}%
\bibitem [{\citenamefont {Ma}\ \emph {et~al.}(2017)\citenamefont {Ma}, \citenamefont {Fang}, \citenamefont {Chen}, \citenamefont {Jin}, \citenamefont {Wu}, \citenamefont {Huang}, \citenamefont {Xia},\ and\ \citenamefont {Tao}}]{ma2017layer}%
  \BibitemOpen
  \bibfield  {author} {\bibinfo {author} {\bibfnamefont {J.}~\bibnamefont {Ma}}, \bibinfo {author} {\bibfnamefont {C.}~\bibnamefont {Fang}}, \bibinfo {author} {\bibfnamefont {C.}~\bibnamefont {Chen}}, \bibinfo {author} {\bibfnamefont {L.}~\bibnamefont {Jin}}, \bibinfo {author} {\bibfnamefont {J.}~\bibnamefont {Wu}}, \bibinfo {author} {\bibfnamefont {H.}~\bibnamefont {Huang}}, \bibinfo {author} {\bibfnamefont {R.}~\bibnamefont {Xia}}, \ and\ \bibinfo {author} {\bibfnamefont {X.}~\bibnamefont {Tao}},\ }\href@noop {} {\bibfield  {journal} {\bibinfo  {journal} {ACS Applied Materials \& Interfaces}\ }\textbf {\bibinfo {volume} {9}},\ \bibinfo {pages} {33925} (\bibinfo {year} {2017})}\BibitemShut {NoStop}%
\bibitem [{\citenamefont {Tan}\ \emph {et~al.}(2016)\citenamefont {Tan}, \citenamefont {Chen},\ and\ \citenamefont {Ghosh}}]{tan2016first}%
  \BibitemOpen
  \bibfield  {author} {\bibinfo {author} {\bibfnamefont {Y.}~\bibnamefont {Tan}}, \bibinfo {author} {\bibfnamefont {F.}~\bibnamefont {Chen}}, \ and\ \bibinfo {author} {\bibfnamefont {A.~W.}\ \bibnamefont {Ghosh}},\ }\href@noop {} {\bibfield  {journal} {\bibinfo  {journal} {Physical Review B}\ }\textbf {\bibinfo {volume} {94}},\ \bibinfo {pages} {214101} (\bibinfo {year} {2016})}\BibitemShut {NoStop}%
\bibitem [{\citenamefont {Yeh}\ \emph {et~al.}(2016)\citenamefont {Yeh}, \citenamefont {Jin}, \citenamefont {Zaki} \emph {et~al.}}]{yeh2018direct}%
  \BibitemOpen
  \bibfield  {author} {\bibinfo {author} {\bibfnamefont {P.-C.}\ \bibnamefont {Yeh}}, \bibinfo {author} {\bibfnamefont {W.}~\bibnamefont {Jin}}, \bibinfo {author} {\bibfnamefont {N.}~\bibnamefont {Zaki}},  \emph {et~al.},\ }\href@noop {} {\bibfield  {journal} {\bibinfo  {journal} {Nano Letters}\ }\textbf {\bibinfo {volume} {16}},\ \bibinfo {pages} {953} (\bibinfo {year} {2016})}\BibitemShut {NoStop}%
\end{thebibliography}

%

\end{document}